\documentstyle[preprint,aps]{revtex}

\begin{document}

%
\def \positronon{\bar\psi}
\def \electronon{\psi}

\def \Wnorm{W}
\def \Wms{{\cal{W}}}
\def \Wbar{{\overline{\Wms}}}

\def \PHI{{\mit\Phi}}

\def \pole{{\mit\Pi}}
\def \POLE#1#2{ {\pole_{#1}}^{#2} }

\def \amp{{\cal A}}

\def \Jhat{\skew5\widehat{J}}
\def \Jms{{\cal{J}}}
\def \Jmshat{{\widehat{\Jms}}}
\def \Jslash{\thinspace{\not{\negthinspace J}}}
\def \Jbar{\skew5\bar{\Jms}}

\def \Ims{{\cal{I}}}
\def \Ibar{\skew3\bar{\Ims}}
\def \Ihat{\skew5\widehat{I}}
\def \Xms{{\cal{X}}}
\def \Yms{{\cal{Y}}}

\def \psihat{\skew2\widehat{\psi}}
\def \psibarhat{\skew2\widehat{\skew2\bar{\psi}}}
\def \psibar{\skew3\bar{\psi}}
\def \Psibar{\bar{\Psi}}
\def \Psihat{\widehat{\Psi}}

\def \GAMMA{{\mit\Gamma}}
\def \GAMMAbar{\bar\GAMMA}

\def \Lam{\Lambda}

%
%
\def \dfour#1{  { {d^4{#1}}\over{(2\pi)^4} }\thinspace  }
\def \dn#1{  { {d^d{#1}}\over{(2\pi)^d} }\thinspace  }

%
%
\def \permsum#1#2{\sum_{{\cal{P}}(#1\ldots #2)}}
\def \cycsum#1#2{\sum_{{\cal{C}}(#1\ldots #2)}}
\def \braket#1#2{ \langle #1 \thinspace\thinspace #2 \rangle }
\def \tekarb#1#2{ {\braket{#1}{#2}}^{*} }
\def \bra#1{ \langle #1 | }
\def \ket#1{ | #1 \rangle }
\def \num{{\cal{N}}}
\def \Trace{\enspace{\rm Tr}\thinspace}

\def \eee{{e}}

\def \eps{\epsilon}
\def \vareps{\varepsilon}
\def \down{{}_\downarrow}
\def \up{{}_\uparrow}
\def \ts{\thinspace}
\newcommand {\beq}{\begin{equation}}
\newcommand {\eeq}{\end{equation}}
\newcommand {\beqa}{\begin{eqnarray}}
\newcommand {\eeqa}{\end{eqnarray}}
\newcommand {\LF}{\nonumber\\}

\def \zee{{\cal{Z}}}
\def \Poff{{\cal{P}}}
\def \Qoff{{\cal{Q}}}
\def \Koff{{\cal{K}}}
\def \half{ \hbox{$1\over2$} }

\def \link#1#2#3{ {{\braket{#1}{#3}}\over{\bra{#1}#2\ket{#3}}} }
\def \invlink#1#2#3{ {{\bra{#1}#2\ket{#3}}\over{\braket{#1}{#3}}} }

\def \linkstar#1#2#3{ { {{\braket{#1}{#3}}^{*}}
\over{{\bra{#1}#2\ket{#3}}^{*}}} }

\def \epsslash{\thinspace{\not{\negthinspace \eps}}}
\def \Poffslash{\thinspace{\not{\negthinspace\negthinspace\Poff}}}

\hyphenation{ap-pen-dix  spin-or  spin-ors}

\draft
\preprint{
Fermilab--Pub--93/327-T
}
\title{One Loop Multiphoton Helicity Amplitudes}
\author{Gregory Mahlon \cite{byline}}
\address{Fermi National Accelerator Laboratory \\
P.O. Box 500 \\
Batavia, IL  60510 }
\date{November 1993}
\maketitle
\begin{abstract}
We use the solutions to
the recursion relations for double-off-shell fermion currents
to compute  helicity amplitudes for $n$-photon scattering
and electron-positron annihilation to photons in the massless
limit of QED.  The form of these solutions is simple enough
to allow {\it all}\ of the integrations to be performed explicitly.
For $n$-photon scattering, we find that
unless $n=4$, the  amplitudes for the helicity
configurations (${+}{+}{+}\cdots{+}$) and (${-}{+}{+}\cdots{+}$)
vanish to one-loop order.

\pacs{11.80.Cr, 12.20.Ds}

\end{abstract}

\section{Introduction}

Much progress has been made during the last decade or so in
the evaluation of tree-level processes containing multiple
gauge bosons, both in QCD and the high energy limit of
the Weinberg-Salam-Glashow model.  Three major
techniques have aided this progress.
First, the multispinor representation
of a gauge field [\ref{SCHWINGER}--\ref{Other}]
allows us to treat the gauge
bosons in a theory on equal footing with the fermions  by
replacing the single Lorentz index on the polarization vectors
with a pair of spinor indices.
With the proper choice of
basis, the gauge boson polarizations  factorize on these
indices, allowing long strings of Dirac matrices to be
written as several short ones.
Second, the color factorization
of QCD amplitudes \cite{colorfactorization}
helps
to organize the many terms in a calculation into gauge-invariant
sub-groups, each proportional to a different color structure.
A final ingredient is the introduction of currents and the
recursion relations that they satisfy [\ref{BGrecursion}--\ref{First}].
The currents are defined to be the sum of all tree graphs containing
exactly $n$ gauge bosons (and possibly a single scalar or spinor
line) with one off-shell particle.  These currents satisfy
relatively simple recursion relations, allowing expressions
involving many gauge bosons to be built from those involving
fewer.  Tree-level amplitudes may be obtained from the currents either
by putting the off-shell particle on shell, or by combining
two or more currents using the vertices of the theory in question.
Of note is the existence of explicit closed-form solutions for
these currents for certain special helicity configurations.

Recently, recursion relations for currents with two off-shell
particles have been obtained
\cite{Dunn,secondpaper,thirdpaper}.
With the presence of a second off-shell particle, the possibility
of forming a one-loop amplitude from tree-level currents
exists.  The purpose of this paper is to pursue that idea
in the case of massless QED.  Application of this method to
QCD will be discussed elsewhere \cite{fifthpaper}.

The processes we will consider are
\beq
\gamma \gamma \rightarrow \gamma\gamma\ldots\gamma
\label{ngamma}
\eeq
and
\beq
e^{+} e^{-} \rightarrow \gamma\gamma\ldots\gamma.
\label{Compton}
\eeq
The Feynman diagrams describing both~(\ref{ngamma})
and~(\ref{Compton})  may be built from a
double-off-shell fermion current:  that is, a fermion line
which radiates $n$ photons all possible ways and  has
both ends off shell.  We have been able to solve the recursion
relation for this current in the case of like-helicity
photons.  As a result, we can compute the one-loop
amplitudes for~(\ref{ngamma}) involving the
helicity configurations $(+++\cdots++)$ and $(-++\cdots++)$.
For~(\ref{Compton}) we are able to obtain amplitudes for
either helicity of the fermion line, with $n$ like-helicity
photons.

The processes listed above
are especially well-suited to be starting points
in an investigation of how to extend the use of recursive
calculations from tree level to one-loop processes.
In the indicated helicity configurations, both processes
vanish at tree level.  As a consequence, both processes must
be ultraviolet and infrared finite at the one-loop level.
Furthermore, it follows from the vanishing of~(\ref{Compton})
at tree level for like-helicity photons, plus the Cutkosky
rules, that these particular one-loop diagrams do not possess
any cuts.  Thus, the results for these diagrams should be
relatively simple.
One might hope that the steps required to reach this simple
endpoint could be made similarly simple,
in contrast to the intermediate steps
involved in a conventional calculation which contain
logarithmic terms that ultimately cancel.
Indeed, we shall see that this is the case.

Our discussion is organized as follows.
In Sec.~II, we  review
the double-off-shell fermion currents  of Ref.~\cite{secondpaper}.
These currents form the basis of the loop amplitudes
presented in this paper.  Of special note is the solution
for the current containing $n$ like-helicity photons.
We examine the one-loop $n$-photon scattering amplitude in
Sec.~III.  Because of the favorable form of the double-off-shell
fermion current appearing in this amplitude, we are able to
evaluate the integrals exactly for arbitrary $n$.
We find the surprising result that
{\it all}\ of the photon-photon scattering amplitudes vanish
for $n\ge5$ in the two helicity configurations we are able
to evaluate.  The one-loop contribution to~(\ref{Compton}) is
the topic of Sec.~IV.  Once again, we find that the relatively
simple form of the double-off-shell fermion current allows us
to perform all of the integrations exactly, producing a
fairly compact result for arbitrary $n$.  We conclude with a
few closing remarks in Sec.~V.

\section{The double-off-shell fermion currents}

In this section we will review the recursion relations
and solutions for the double-off-shell fermion currents
presented in Ref.~\cite{secondpaper}.
A summary of our conventions and notation for the Weyl--van der
Waerden spinors used in this discussion may be found
in Ref.~\cite{firstpaper}.

We define the $n$-particle double off-shell
fermion current to consist of the sum of
all tree graphs containing exactly $n$ photons attached to
a single fermion line all possible ways.  Both ends of
the fermion line are off shell.  All momenta are directed inward.
We will denote the momenta of the photons by $k_1, k_2, \ldots,
k_n$.  The off-shell positron has momentum $\Poff$, while the
momentum of the off-shell electron is $\Qoff$.  Momentum conservation
relates the momenta via
\beq
\Poff + k_1 + k_2 + \cdots + k_n + \Qoff \equiv
\Poff + \kappa(1,n) + \Qoff = 0
\label{momcon}
\eeq
Since the helicity of the fermion is conserved in the massless limit,
we have two different (but related) double off-shell fermion currents.
Let us denote the left-handed current by
$\Psi_{\alpha\dot\alpha}(\Poff;1,\ldots,n;\Qoff)$ and
the right-handed current by
$\Psibar^{\dot\alpha\alpha}(\Poff;1,\ldots,n;\Qoff)$.
Because of~(\ref{momcon}), the
argument lists are overspecified.  When convenient, we will suppress
either $\Poff$ or $\Qoff$.
Because of the way we have defined these currents, the order of the
photon arguments is irrelevant.

In Ref. \cite{secondpaper} we find that the left-handed
double-off-shell fermion current satisfies the following recursion
relation:
\beq
\Psi_{\alpha\dot\alpha}(\Poff;1,\ldots,n) = -e\sqrt2
\permsum{1}{n}
{
{1}
\over
{(n{-}1)!}
}
\Psi_{\alpha\dot\beta}(\Poff;1,\ldots,n{-}1)
\bar\eps^{\dot\beta\beta}(n)
{
{ [\Poff{+}\kappa(1,n)]_{\beta\dot\alpha} }
\over
[\Poff{+}\kappa(1,n)]^2
} .
\label{PSIrec}
\eeq
The notation ${\cal{P}}(1\cdots n)$ indicates a  sum
over all permutations of the $n$ photons.
The zero-photon current is just a propagator for the fermion:
\beq
\Psi_{\alpha\dot\alpha}(\Poff;\Qoff) =
{
{-i \Poff_{\alpha\dot\alpha} }
\over
{ \Poff^2 }
}
=
{
{i \Qoff_{\alpha\dot\alpha} }
\over
{ \Qoff^2 }
}
\label{PSIzero}
\eeq
(This is one expression in which it is important to include both
$\Poff$ and $\Qoff$ in the argument list to avoid ambiguity.)

The right-handed current  satisfies the recursion relation
\beq
\Psibar^{\dot\alpha\alpha}(\Poff;1,\ldots,n) = -e\sqrt2
\permsum{1}{n}
{
{1}
\over
{(n{-}1)!}
}
\Psibar^{\dot\alpha\beta}(\Poff;1,\ldots,n{-}1)
\eps_{\beta\dot\beta}(n)
{
{ [\bar\Poff{+}\bar\kappa(1,n)]^{\dot\beta\alpha} }
\over
[\Poff{+}\kappa(1,n)]^2
} .
\label{PSIbarrec}
\eeq
It is closely related to the left-handed current via
the crossing relation
\beq
\Psibar^{\dot\alpha\alpha}(\Poff;1,\ldots,n;\Qoff) =
(-1)^{n+1}
\vareps^{\alpha\beta} \vareps^{\dot\alpha\dot\beta}
\Psi_{\beta\dot\beta}(\Qoff;1,\ldots,n;\Poff).
\label{PSIcross}
\eeq

In Ref.~\cite{secondpaper}, we were able to obtain a solution for the
left-handed current in the case where all of
the photons have the same helicity.  In this situation, the photon
polarizations may be written as~\cite{BG}
\beq
\eps_{\alpha\dot\alpha}(j^{+}) =
{
{ u_{\alpha}(g) \bar u_{\dot\alpha}(k_j) }
\over
{ \braket{k}{g} }
}
\label{allpluspol}
\eeq
for the $j$th photon.  In Eq.~(\ref{allpluspol}) $g$ is an arbitrary
null momentum, partially reflecting
the gauge freedom associated with
QED.  The choice of $g$ does not
affect any physics result.  In general,
a different value of $g$ may be chosen
for each photon; however, the choice of Eq.~(\ref{allpluspol})
leads to  the useful property
\beq
\bar\eps^{\dot\alpha\alpha}(j^{+})
\eps_{\alpha\dot\alpha}(\ell^{+}) = 0
\label{killer}
\eeq
for any pair of positive helicity polarization spinors.

The solution to the recursion relation given in Eq.~(\ref{PSIrec})
using the gauge choice~(\ref{allpluspol})
was found in Ref. \cite{secondpaper} to be
\beqa
u^{\alpha}(g) \Psi_{\alpha\dot\alpha}&& (\Poff;1^{+},\ldots,n^{+})
= \LF
&&-i(-e\sqrt2)^n
\permsum{1}{n}
{
{ u^{\alpha}(g) [\Poff + \kappa(1,n)]_{\alpha\dot\alpha} }
\over
{ \bra{g} 1,\ldots,n \ket{g} }
}
\sum_{j=1}^n
u^{\beta}(g) \POLE{\beta}{\gamma} (\Poff,1,\ldots,j) u_{\gamma}(g),
\label{LHsolution}
\eeqa
where
\beq
\POLE{\beta}{\gamma}(\Poff,1,\ldots,j)
\equiv
{
{ k_{j\beta\dot\beta}[\bar\Poff+\bar\kappa(1,j{-}1)]^{\dot\beta\gamma} }
\over
{ [\Poff + \kappa(1,j{-}1)]^2 [\Poff + \kappa(1,j)]^2 }
}.
\label{poledef}
\eeq
Note that the zero-photon current~(\ref{PSIzero}) does not
fit into the form~(\ref{LHsolution}).  Also,
although the combination
$u^{\alpha}(g) \Psi_{\alpha\dot\alpha}$
takes a convenient  form, we have not
been able to find a similarly
compact expression for $\Psi_{\alpha\dot\alpha}$ itself.  Usually,
this is not a serious problem.

{}From the left-handed current given in~(\ref{LHsolution}) and
the crossing relation~(\ref{PSIcross}), it is not hard to
see that
\beqa
\Psibar^{\dot\alpha\alpha} (\Poff;1^{+}&&,\ldots,n^{+})
u_{\alpha}(g)
=  \LF &&
-i(-e\sqrt2)^n
\permsum{1}{n}
{
{ \bar\Poff^{\dot\alpha\alpha}u_{\alpha}(g) }
\over
{ \bra{g} 1,\ldots,n \ket{g} }
}
\sum_{j=1}^n
u^{\beta}(g) \POLE{\beta}{\gamma} (\Poff,1,\ldots,j) u_{\gamma}(g),
\label{RHsolution}
\eeqa

\section{Photon-photon scattering}\label{photonphoton}

We begin the application of the double-off-shell current
to loop processes with a discussion of photon-photon scattering.
Photon-photon interactions have long been of theoretical
interest,
the first complete computation  for four photons
being performed by Karplus and Neuman \cite{KarplusNeuman}.
Helicity-projected amplitudes, including finite mass effects for
the electron, were first obtained by
Costantini, De Tollis, and Pistoni \cite{Costantini}.
Gastmans and Wu \cite{ubiquitous} use this process to
illustrate the use of spinor helicity methods at loop
level, presenting the limit where the electron mass
may be ignored.
In this section, we will not only reproduce the massless limit
results, but we will present  amplitudes
involving more than four photons, albeit for a restricted
set of helicity configurations.

\subsection{Preliminaries}

Figure~\ref{ngammafig} illustrates how to utilize the
double-off-shell fermion current
in the amplitude for the self-interaction
of $n$ photons.
If we sum over all permutations of the $n$ photons, we overcount
by a factor of $n$ (the $n$ cyclic permutations of the
photons do not produce distinct Feynman diagrams, as may be seen
by shifting the loop momentum).
For our purposes,
it is convenient
circumvent this problem by excluding the $n$th photon from
the permutation sum, rather than including it and dividing
by $n$.
Applying the QED Feynman rules
to Fig.~\ref{ngammafig}
we obtain
\beq
\amp(1,\ldots,n) =
- \int  \dfour{\Poff}
\Lam(1,\ldots,n)
\label{ngammaintegral}
\eeq
where
\beq
\Lam(1,\ldots,n) = -ie \Trace[\epsslash(n) \Psi(\Poff;1,\ldots,n{-}1)].
\label{lamdef}
\eeq
In order to write~(\ref{lamdef}) in spinor form, we break the
trace into left- and right-handed contributions
by inserting $1 = \half (1+\gamma_5) + \half (1-\gamma_5)$.
Thus, we obtain
\beq
\Lam(1,\ldots,n) =
-ie \sqrt2 \thinspace
\bigl[
\bar\eps^{\dot\alpha\alpha}(n)
\Psi_{\alpha\dot\alpha}(\Poff;1,\ldots,n{-}1)
+\eps_{\alpha\dot\alpha}(n)
\Psibar^{\dot\alpha\alpha}(\Poff;1,\ldots,n{-}1)
\bigr].
\label{lamspinor}
\eeq

Equation~(\ref{lamspinor}) is valid for any combination
of photon helicities.  We need only know the expressions
for the currents appearing on the right hand side.
We now focus on the two amplitudes that our knowledge
of $\Psi(\Poff;1^{+},\ldots,n^{+})$ allows us to obtain.
Our first option is to use Eq.~(\ref{allpluspol}) for
all of the photons and so  compute $\amp(1^{+},\ldots,n^{+})$.
Note that this gauge choice leaves $g$ as an arbitrary parameter
which must cancel in the final result.  Our other option is
to write
\beq
\eps_{\alpha\dot\alpha}(n^{-}) =
{
{ u_{\alpha}(k_n) \bar u_{\dot\alpha}(h) }
\over
{ {\braket{n}{h}}^{*} }
}
\label{neghel}
\eeq
for the $n$th photon, and set $g=k_n$
in the other polarizations~\cite{BG}.  This gives us access to
$\amp\bigl(1^{+},\ldots,(n{-}1)^{+},n^{-}\bigr)$.
The indicated gauge  choice
has the virtue of satisfying not only~(\ref{killer}),
but also
\beq
\bar\eps^{\dot\alpha\alpha}(j^{+})
\eps_{\alpha\dot\alpha}(n^{-}) = 0.
\label{negkiller}
\eeq
The arbitrary null vector $h$ should not appear
in the final result.  We will concentrate our discussion on
$\amp(1^{+},\ldots,n^{+})$. The computation of
$\amp\bigl(1^{+},\ldots,(n{-}1)^{+},n^{-}\bigr)$ proceeds
in essentially the same manner.

\subsection{The momentum integral}  \label{INTEGRATION}

It would seem that all we have to do at this stage is
to insert the expressions for the polarizations and the currents and
perform the momentum integration.
However, this is not quite correct.
A na{\"\i}ve application of
the solutions to the recursion relations
to~(\ref{lamspinor}) yields
\beq
\Lam(1^{+},\ldots,n^{+}) =
(-e\sqrt2)^n \negthinspace
\permsum{1}{n-1}
{
{  2\bar u_{\dot\alpha}(k_n)\bar\Poff^{\dot\alpha\alpha} u_{\alpha}(g) }
\over
{ \bra{g} 1,\ldots,n{-}1 \ket{g} \braket{n}{g} }
}
\sum_{j=1}^{n-1}
{
{ u^{\beta}(g)k_{j\beta\dot\beta}
  [\bar\Poff{+}\bar\kappa(1,j)]^{\dot\beta\gamma}
  u_{\gamma}(g) }
\over
{ [\Poff{+}\kappa(1,j{-}1)]^2 [\Poff{+}\kappa(1,j)]^2 }
},
\label{wrong}
\eeq
the factor of 2 resulting from equal contributions by the left- and
right-handed fermion lines.
If this expression were inserted into (\ref{ngammaintegral}),
power counting would indicate a superficial quadratic divergence.
Actually, the divergence is only logarithmic, as the coefficient of
the highest power of $\Poff$ is proportional to $\braket{g}{g}$,
which vanishes.
Since it is well-known that the amplitude
for light-by-light scattering is
finite,
any divergences apparently present were introduced in
the intermediate steps.
In order to cancel these divergences and obtain the
correct finite result, we must first examine their origin.
The process of reducing the double-off-shell
current into the form given in~(\ref{LHsolution}) involves repeated use
of identities such as
\beq
k_j = [\Poff + \kappa(1,j)] - [\Poff + \kappa(1,j{-}1)]
\label{FeynmanID}
\eeq
to convert the expression inside the permutation sum from one
with  a single
term and $n$ propagators into one with
$n$ terms and two propagators.  The two terms on the right hand side
of~(\ref{FeynmanID}) each contain one more power of $\Poff$ than the
one on the right hand side.  If the integral corresponding to the
left hand side of~(\ref{FeynmanID}) is finite, then the combination
of the two integrals on the right is also finite.
However,
it is possible that both  terms on the right diverge
when considered separately, a finite quantity being
obtained only from their combination.
In that case,
to obtain the correct result from the right hand side,
we must
treat the two pieces in an identical manner.
In particular, shifts of the integration momentum in one term
relative to the other term are forbidden.
Because of the
way the ``fragments'' from~(\ref{FeynmanID}) recombine to form the
final result~(\ref{LHsolution}) [see the discussion
following Eq.~(\ref{almostreduced})], a straightforward integration
of~(\ref{wrong}) involves such forbidden shifts.
Thus, simply regulating the integral obtained from~(\ref{wrong})
does not give the correct result.

Instead, we must turn to the recursion relation to let us ``back-up''
one stage in the reduction.  That is, we use~(\ref{PSIrec})
and~(\ref{PSIbarrec}) to write the
integrand~(\ref{lamspinor}) in terms of
the $(n{-}2)$--photon current instead of the $(n{-}1)$--photon current.
The result of this procedure is
\beqa
\Lam(1^{+},\ldots,n^{+}) =
(-e\sqrt2)^n
\permsum{1}{n-1}
&&
{
{ 2\bar u_{\dot\alpha}(k_n) \bar\Poff^{\dot\alpha\alpha} u_{\alpha}(g) }
\over
{ \bra{g} 1,\ldots,n{-}2 \ket{g}  \braket{n}{g} }
}
{
{  \bar u_{\dot\delta}(k_{n-1})
   [ \bar\Poff + \bar\kappa(1,n{-}1) ]^{\dot\delta\delta}
   u_{\delta}(g) }
\over
{ \braket{n{-}1}{g} [\Poff + \kappa(1,n{-}1)]^2  }
}
\LF && \times
\sum_{j=1}^{n-2}
{
{ u^{\beta}(g)k_{j\beta\dot\beta}
  [\bar\Poff{+}\bar\kappa(1,j)]^{\dot\beta\gamma}
  u_{\gamma}(g) }
\over
{ [\Poff{+}\kappa(1,j{-}1)]^2 [\Poff{+}\kappa(1,j)]^2 }
}.
\label{corrected}
\eeqa
Power counting of Eq.~(\ref{corrected}) implies a possible linear
divergence; however, the coefficients of both
the linearly and logarithmically divergent pieces are proportional
to $\braket{g}{g}$ and thus vanish.  The expression is actually
convergent.

Since~(\ref{corrected}) converges, we may impose a regulator on it, and
then carry out the reduction to two denominators on the regulated
expression.  We should like to employ dimensional regularization,
and so must consider how
to continue~(\ref{corrected}) to $d$ dimensions.
Fortunately, it is sufficient to extend only the internal momentum
$\Poff$, leaving the external momenta and polarization vectors in
4 dimensions \cite{Regulator}.
It is not possible to translate an object
like $\bar\Poff^{\dot\alpha\alpha}$
into $d$ dimensions, as it
corresponds to
$\half(1{+}\gamma_5)\Poffslash\half(1{-}\gamma_5)$.  However,
note that every occurrence of $\bar\Poff^{\dot\alpha\alpha}$
in~(\ref{corrected}) may
be rewritten as a Lorentz
dot product with a polarization, forming $\Poff\cdot\eps(\ell^{+})$
with various values of $\ell$.
We may extend this form to $d$ dimensions.

To facilitate a quick return
to multispinor notation, we
decompose the $n$-dimensional vector $\Poff$ into a 4-dimensional
piece $P$ and a ($n{-}4$)-dimensional piece $m$~\cite{SPLITUP},
\beq
\Poff = P + m.
\eeq
Only
the usual 4 space-time components of $P$ are non-zero, while
only the ``extra'' components of $m$ are non-zero.
Hence,
$m$ dotted into any 4 dimensional vector vanishes.
We set  $ m^2 \equiv -\mu^2$, and adopt $\mu$ as the
radial integration variable in the ($n{-}4$)-dimensional subspace.
The integration measure appearing
in~(\ref{ngammaintegral}) is replaced by
\beq
\int \dn{\Poff} = \int \dfour{P}
{
{ -\eps (4\pi)^{\eps} }
\over
{ \Gamma(1-\eps) }
}
\int_{0}^{\infty}
d\mu^2 \enspace  (\mu^2)^{-1-\eps}.
\label{newmeasure}
\eeq

Since every occurrence of $\Poff$ in the numerator of~(\ref{corrected})
is as $\Poff\cdot\eps(\ell^{+})$, the continuation to $d$ dimensions
simply  implies
\beq
\Poff \cdot \eps(\ell^{+}) \rightarrow P \cdot \eps(\ell^{+}).
\eeq
Furthermore, the
denominators simply pick up an extra term, $-\mu^2$.
Thus, we obtain
\beqa
\Lam(1^{+},\ldots,n^{+}) =
(-e\sqrt2)^n \negthinspace
\permsum{1}{n-1}
&&
{
{  2 \bar u_{\dot\alpha}(k_n) \bar P^{\dot\alpha\alpha} u_{\alpha}(g) }
\over
{ \bra{g} 1,\ldots,n{-}2 \ket{g}  \braket{n}{g} }
}
{
{  \bar u_{\dot\delta}(k_{n-1})
   [ \bar P + \bar\kappa(1,n{-}1) ]^{\dot\delta\delta}
   u_{\delta}(g) }
\over
{ \braket{n{-}1}{g}\{ [P + \kappa(1,n{-}1)]^2{-}\mu^2\}  }
}
\LF && \times
\sum_{j=1}^{n-2}
{
{ u^{\beta}(g)k_{j\beta\dot\beta}
  [\bar P{+}\bar\kappa(1,j)]^{\dot\beta\gamma}
  u_{\gamma}(g) }
\over
{ \{[P{+}\kappa(1,j{-}1)]^2{-}\mu^2\} \{[P{+}\kappa(1,j)]^2{-}\mu^2\} }
}.
\label{regulated}
\eeqa
We now attempt to reduce~(\ref{regulated}) to a form containing just
two propagators, like~(\ref{wrong}).

We begin by multiplying
by $\braket{n{-}2}{n{-}1} / \braket{n{-}2}{n{-}1}$ and
writing
\beq
\braket{n{-}2}{n{-}1}\braket{g}{j} =
\braket{n{-}2}{g}\braket{n{-}1}{j} - \braket{n{-}1}{g}\braket{n{-}2}{j}
\eeq
(the Schouten identity) to obtain
\beqa
\Lam(1^{+},&&\ldots,n^{+}) =
\LF &&
2(-e\sqrt2)^n \negthinspace
\permsum{1}{n-1}
{
{   \bar u_{\dot\alpha}(k_n) \bar P^{\dot\alpha\alpha} u_{\alpha}(g) }
\over
{ \bra{g} 1,\ldots,n{-}1 \ket{g}  \braket{n}{g} }
}
\LF && \qquad\qquad\qquad \times
\sum_{j=1}^{n-2}
{
{ u^{\delta}(g)
  [P{+}\kappa(1,n{-}1)]_{\delta\dot\delta}
  \bar k_{n-1}^{\dot\delta\beta}
  k_{j\beta\dot\beta}
  [\bar P{+}\bar\kappa(1,j)]^{\dot\beta\gamma}
  u_{\gamma}(g) }
\over
{ \{[P + \kappa(1,n{-}1)]^2{-}\mu^2\}
  \{[P{+}\kappa(1,j{-}1)]^2{-}\mu^2\} \{[P{+}\kappa(1,j)]^2{-}\mu^2\} }
} \LF
-&& 2(-e\sqrt2)^n \negthinspace
\permsum{1}{n-1}
{
{  \bar u_{\dot\alpha}(k_n) \bar P^{\dot\alpha\alpha} u_{\alpha}(g) }
\over
{ \bra{g} 1,\ldots,n{-}2 \ket{g}  \braket{n}{g} }
}
{
{  \bar u_{\dot\delta}(k_{n-1})
   [ \bar P + \bar\kappa(1,n{-}1) ]^{\dot\delta\delta}
   u_{\delta}(g) }
\over
{ \{[P + \kappa(1,n{-}1)]^2{-}\mu^2\}  }
}
\LF && \qquad\qquad\qquad \times
\sum_{j=1}^{n-3}
{
{ \braket{n{-}2}{j} }
\over
{ \braket{n{-}2}{n{-}1} }
}
{
{ \bar u_{\dot\beta}(k_j)
  [\bar P{+}\bar\kappa(1,j)]^{\dot\beta\gamma}
  u_{\gamma}(g) }
\over
{ \{[P{+}\kappa(1,j{-}1)]^2{-}\mu^2\} \{[P{+}\kappa(1,j)]^2{-}\mu^2\} }
}.
\label{Fierzed}
\eeqa
We now focus on the following portion of the first term:
\beq
\num_1 \equiv
u^{\delta}(g)
[P{+}\kappa(1,n{-}1)]_{\delta\dot\delta}
\bar k_{n-1}^{\dot\delta\beta}
k_{j\beta\dot\beta}
[\bar P{+}\bar\kappa(1,j)]^{\dot\beta\gamma}
u_{\gamma}(g).
\label{numdef}
\eeq
By clever application of identities similar to~(\ref{FeynmanID}),
we may rewrite $\num_1$ as
\beqa
\num_1 = &\enspace&
[P+\kappa(1,n{-}1)]^2 \thinspace
u^{\delta}(g)
k_{j\delta\dot\beta}
[\bar P + \bar\kappa(1,j)]^{\dot\beta\gamma}
u_{\gamma}(g)
\LF & - &
[P+\kappa(1,j)]^2 \thinspace
u^{\delta}(g)
[P+\kappa(1,n{-}1)]_{\delta\dot\delta}
[\bar P + \bar\kappa(1,j{-}1)]^{\dot\delta\gamma}
u_{\gamma}(g)
\LF & + &
[P+\kappa(1,j{-}1)]^2 \thinspace
u^{\delta}(g)
[P+\kappa(1,n{-}1)]_{\delta\dot\delta}
[\bar P + \bar\kappa(1,j)]^{\dot\delta\beta}
u_{\beta}(g)
\LF & - &
u^{\delta}(g)
[P+\kappa(1,n{-}1)]_{\delta\dot\delta}
\bar\kappa^{\dot\delta\beta}(j{+}1,n{-}2)
k_{j\beta\dot\beta}
[\bar P + \bar\kappa(1,j)]^{\dot\beta\gamma}
u_{\gamma}(g).
\label{splitup}
\eeqa
If the regulator were not present, the quadratic factors
appearing in the first three terms of~(\ref{splitup})
would each produce terms containing only two propagators.
Since this is precisely what we want to occur,
we add and subtract the appropriate terms.
Thus, when we combine~(\ref{splitup}) with~(\ref{Fierzed})
we obtain
\beqa
\Lam(1^{+},&&\ldots,n^{+}) =
\LF &&
2(-e\sqrt2)^n \negthinspace
\permsum{1}{n-1}
{
{   \bar u_{\dot\alpha}(k_n) \bar P^{\dot\alpha\alpha} u_{\alpha}(g) }
\over
{ \bra{g} 1,\ldots,n{-}1 \ket{g}  \braket{n}{g} }
}
\LF && \qquad\qquad\qquad \times
\sum_{j=1}^{n-2}
{
{ u^{\delta}(g)
  k_{j\delta\dot\beta}
  [\bar P{+}\bar\kappa(1,j)]^{\dot\beta\gamma}
  u_{\gamma}(g) }
\over
{ \{[P{+}\kappa(1,j{-}1)]^2{-}\mu^2\} \{[P{+}\kappa(1,j)]^2{-}\mu^2\} }
} \LF
-&& 2(-e\sqrt2)^n \negthinspace
\permsum{1}{n-1}
{
{   \bar u_{\dot\alpha}(k_n) \bar P^{\dot\alpha\alpha} u_{\alpha}(g) }
\over
{ \bra{g} 1,\ldots,n{-}1 \ket{g}  \braket{n}{g} }
}
{
{ u^{\delta}(g)[P{+}\kappa(1,n{-}1)]_{\delta\dot\delta} }
\over
{ \{[P{+}\kappa(1,n{-}1)]^2-\mu^2\} }
}
\LF && \qquad\qquad\qquad \times
\sum_{j=1}^{n-2}
\Biggl\{
{
{ [\bar P{+}\bar\kappa(1,j{-}1)]^{\dot\delta\gamma}
  u_{\gamma}(g) }
\over
{ \{[P{+}\kappa(1,j{-}1)]^2{-}\mu^2\} }
}  -
{
{ [\bar P{+}\bar\kappa(1,j)]^{\dot\delta\beta}
  u_{\beta}(g) }
\over
{ \{[P{+}\kappa(1,j)]^2{-}\mu^2\} }
}
\Biggr\}
\LF
-&& 2(-e\sqrt2)^n  \negthinspace
\permsum{1}{n-1}
{
{   \bar u_{\dot\alpha}(k_n) \bar P^{\dot\alpha\alpha} u_{\alpha}(g) }
\over
{ \bra{g} 1,\ldots,n{-}1 \ket{g}  \braket{n}{g} }
}
{
{ u^{\delta}(g) [P{+}\kappa(1,n{-}1)]_{\delta\dot\delta} }
\over
{ \{[P + \kappa(1,n{-}1)]^2{-}\mu^2\} }
}
\LF && \qquad\qquad\qquad \times
\sum_{j=1}^{n-3}
{
{ \bar \kappa^{\dot\delta\beta}(j{+}1,n{-}2)
  k_{j\beta\dot\beta}
  [\bar P{+}\bar\kappa(1,j)]^{\dot\beta\gamma}
  u_{\gamma}(g) }
\over
{ \{[P{+}\kappa(1,j{-}1)]^2{-}\mu^2\} \{[P{+}\kappa(1,j)]^2{-}\mu^2\} }
} \LF
-&& 2(-e\sqrt2)^n  \negthinspace
\permsum{1}{n-1}
{
{  \bar u_{\dot\alpha}(k_n) \bar P^{\dot\alpha\alpha} u_{\alpha}(g) }
\over
{ \bra{g} 1,\ldots,n{-}2 \ket{g}  \braket{n}{g} }
}
{
{  \bar u_{\dot\delta}(k_{n-1})
   [ \bar P + \bar\kappa(1,n{-}1) ]^{\dot\delta\delta}
   u_{\delta}(g) }
\over
{ \{[P + \kappa(1,n{-}1)]^2{-}\mu^2\}  }
}
\LF && \qquad\qquad\qquad \times
\sum_{j=1}^{n-3}
{
{ \braket{n{-}2}{j} }
\over
{ \braket{n{-}2}{n{-}1} }
}
{
{ \bar u_{\dot\beta}(k_j)
  [\bar P{+}\bar\kappa(1,j)]^{\dot\beta\gamma}
  u_{\gamma}(g) }
\over
{ \{[P{+}\kappa(1,j{-}1)]^2{-}\mu^2\} \{[P{+}\kappa(1,j)]^2{-}\mu^2\} }
} \LF
-&& 2\mu^2(-e\sqrt2)^n  \negthinspace
\permsum{1}{n-1}
{
{   \bar u_{\dot\alpha}(k_n) \bar P^{\dot\alpha\alpha} u_{\alpha}(g) }
\over
{ \bra{g} 1,\ldots,n{-}1 \ket{g}  \braket{n}{g} }
}
{
{1}
\over
{ \{[P + \kappa(1,n{-}1)]^2{-}\mu^2\} }
}
\LF && \quad\qquad\qquad\qquad \times
\sum_{j=1}^{n-2}
{
{ u^{\delta}(g)
  k_{j\delta\dot\gamma}
  \bar\kappa^{\dot\gamma\gamma}(j{+}1,n{-}1)
  u_{\gamma}(g) }
\over
{ \{[P{+}\kappa(1,j{-}1)]^2{-}\mu^2\} \{[P{+}\kappa(1,j)]^2{-}\mu^2\} }
}.
\label{almostreduced}
\eeqa
The sum on $j$
appearing in the second term of~(\ref{almostreduced})
may be performed trivially, with only the endpoint terms surviving.
One of the resultant terms vanishes when the permutation
sum is performed; the other term may be  used to extend the
sum appearing in
the first term to include $j=n{-}1$.
This converts the first term into precisely the result that
we would have obtained by regulating~(\ref{wrong}) directly.
The third and fourth
terms of~(\ref{almostreduced}) may be shown to cancel by
writing out the implicit sum $\kappa(j{+}1,n{-}2)
= k_{j+1} + \cdots + k_{n-2}$ and
using the permutation sum to relabel the successive terms
thus generated \cite{firstpaper}.
The fifth term does not combine with anything else:  it represents
the effect of imposing the regulator prior to reducing the
integrand.

Thus, the integrals we must consider are
\beqa
\amp_1 \equiv -2(-e\sqrt2)^n  \negthinspace
\permsum{1}{n-1}
\sum_{j=1}^{n-1}
\int
\dn{\Poff}  &&
{
{   \bar u_{\dot\alpha}(k_n) \bar P^{\dot\alpha\alpha} u_{\alpha}(g) }
\over
{ \bra{g} 1,\ldots,n{-}1 \ket{g}  \braket{n}{g} }
}
\LF && \times
{
{ u^{\delta}(g)
  k_{j\delta\dot\beta}
  [\bar P{+}\bar\kappa(1,j)]^{\dot\beta\gamma}
  u_{\gamma}(g) }
\over
{ \{[P{+}\kappa(1,j{-}1)]^2{-}\mu^2\} \{[P{+}\kappa(1,j)]^2{-}\mu^2\} }
},
\label{zerointegral}
\eeqa
generated from the first two terms of~(\ref{almostreduced}),
and
\beqa
\amp_2 \equiv 2(-e\sqrt2)^n  \negthinspace
\permsum{1}{n-1}
\sum_{j=1}^{n-2}
\int
\dn{\Poff}  &&
{
{   \bar u_{\dot\alpha}(k_n) \bar P^{\dot\alpha\alpha} u_{\alpha}(g) }
\over
{ \bra{g} 1,\ldots,n{-}1 \ket{g}  \braket{n}{g} }
}
{
{\mu^2}
\over
{ \{[P + \kappa(1,n{-}1)]^2{-}\mu^2\} }
}
\LF && \times
{
{ u^{\delta}(g)
  k_{j\delta\dot\gamma}
  \bar\kappa^{\dot\gamma\gamma}(j{+}1,n{-}1)
  u_{\gamma}(g) }
\over
{ \{[P{+}\kappa(1,j{-}1)]^2{-}\mu^2\} \{[P{+}\kappa(1,j)]^2{-}\mu^2\} }
},
\label{goodstuff}
\eeqa
from the last term of~(\ref{almostreduced}).

We will begin with $\amp_1$ since it is the simplest.
A single Feynman parameter is sufficient to combine the
denominators, producing
\beq
\Delta_1 = (1-x)\{[P+\kappa(1,j{-}1)]^2-\mu^2\}
+ x \{[P+\kappa(1,j)]^2-\mu^2\}.
\eeq
Expanding and rearranging this expression, we find that
\beq
\Delta_1 =
[P + \kappa(1,j{-}1) + xk_j]^2 - \mu^2.
\label{firstshift}
\eeq
We now apply the momentum shift implied by~(\ref{firstshift})
to the numerator
of~(\ref{zerointegral}).  Since
the combination
\beqa
u^{\delta}(g)
k_{j\delta\dot\beta}
[\bar P{+}\bar\kappa(1,j)]^{\dot\beta\gamma}
u_{\gamma}(g)
\enspace\rightarrow\enspace &&
(1-x) u^{\delta}(g)
k_{j\delta\dot\beta}
k_j^{\dot\beta\gamma}
u_{\gamma}(g) \LF
=&& (1-x) \thinspace k_j^2 \thinspace\braket{g}{g}
= 0,
\label{zeroshift}
\eeqa
appears, we conclude that
the first integral vanishes.

We now turn to the second integral, which contains the entire result.
Introducing three Feynman parameters produces the denominator
\beq
\Delta_2 =
x\{[P{+}\kappa(1,j{-}1)]^2{-}\mu^2\}
+y\{[P{+}\kappa(1,j)]^2{-}\mu^2\}
+z\{[P{+}\kappa(1,n{-}1)]^2{-}\mu^2\} ,
\eeq
which may be rewritten as
\beq
\Delta_2 =
[P + (1{-}z)\kappa(1,j{-}1) + yk_j - zk_n]^2
+ K^2 - \mu^2
\label{secondshift}
\eeq
where
\beq
K^2 \equiv z(1-z) [\kappa(1,j{-}1)+k_n]^2
+2yzk_j\cdot[\kappa(1,j{-}1)+k_n].
\eeq
Shifting the momentum in the manner implied by~(\ref{secondshift})
and doing the integral over the four space-time dimensions yields
\beqa
\amp(1^{+},\ldots,n^{+}) = &&
\LF
{
{i}
\over
{8\pi^2}
}
(-e\sqrt2)^n &&
\negthinspace\negthinspace
\permsum{1}{n-1}
\sum_{j=1}^{n-2}
{
{ u^{\delta}(g)
  k_{j\delta\dot\gamma}
  \bar\kappa^{\dot\gamma\gamma}(j{+}1,n{-}1)
  u_{\gamma}(g) }
\over
{ \bra{g} 1,\ldots,n{-}1 \ket{g}  \braket{n}{g} }
}
\LF && \times
\int_{0}^{\infty} \negthinspace dx
\int_{0}^{\infty} \negthinspace dy
\int_{0}^{\infty} \negthinspace dz
\enspace\delta(1{-}x{-}y{-}z) \thinspace
{ \bar u_{\dot\alpha}(k_n)
 [(1{-}z)\bar\kappa(1,j{-}1)+y\bar k_j]^{\dot\alpha\alpha}
u_{\alpha}(g) }
\LF && \times
{
{ \eps (4\pi)^{\eps} }
\over
{ \Gamma(1-\eps) }
}
\int_{0}^{\infty}
d\mu^2 \enspace
{
{ (\mu^2)^{-\eps}}
\over
{ K^2 - \mu^2 }
}.
\label{Pintdone}
\eeqa
Because an explicit factor of $\eps$ appears in the numerator
of~(\ref{Pintdone}), any pieces
of the integral over $\mu^2$ which are
finite as $\eps\rightarrow0$ are irrelevant.
Thus, we write
\beq
\int_{0}^{\infty}
d\mu^2
{
{ (\mu^2)^{-\eps}}
\over
{ K^2 - \mu^2 }
}
= - { {1}\over{\eps} } + {\cal O}(1).
\label{muint}
\eeq
Inserting~(\ref{muint}) into~(\ref{Pintdone}) and performing
the now trivial Feynman parameter integrals yields
\beqa
\amp(1^{+},\ldots,n^{+}) =
{
{i}
\over
{48\pi^2}
}
(-e\sqrt2)^n \negthinspace\negthinspace
\permsum{1}{n-1} &&
\sum_{j=1}^{n-2}
{
{ u^{\delta}(g)
  k_{j\delta\dot\gamma}
  \bar\kappa^{\dot\gamma\gamma}(j{+}1,n{-}1)
  u_{\gamma}(g) }
\over
{ \bra{g} 1,\ldots,n{-}1 \ket{g}  \braket{n}{g} }
}
\LF &&
\phantom{\biggl[} \times\thinspace
{ \bar u_{\dot\alpha}(k_n)
 [\bar\kappa(j{+}1,n{-}1) - \bar\kappa(1,j{-}1)]^{\dot\alpha\alpha}
u_{\alpha}(g) }.
\label{integrated}
\eeqa
All that remains at this stage is to eliminate the occurrences
of the gauge momentum $g$.

\subsection{Gauge invariance of the result}

We begin by multiplying the expression in~(\ref{integrated})
by $\braket{n{-}1}{1} / \braket{n{-}1}{1}$ and applying
the Schouten identity to obtain
\beqa
\amp(1^{+},\ldots,n^{+}) = &&
{
{i}
\over
{48\pi^2}
}
(-e\sqrt2)^n \negthinspace\negthinspace
\permsum{1}{n-1}
\sum_{j=1}^{n-3}
{
{ u^{\delta}(g)
  k_{j\delta\dot\gamma}
  \bar\kappa^{\dot\gamma\gamma}(j{+}1,n{-}2)
  u_{\gamma}(k_{n-1}) }
\over
{ \bra{n{-}1} 1,\ldots,n{-}1 \ket{g}  \braket{n}{g} }
}
\LF && \qquad\qquad\qquad
\phantom{\biggl[} \times\thinspace
{ \bar u_{\dot\alpha}(k_n)
 [\bar\kappa(j{+}1,n{-}1) - \bar\kappa(1,j{-}1)]^{\dot\alpha\alpha}
u_{\alpha}(g) }
\LF  + &&
{
{i}
\over
{48\pi^2}
}
(-e\sqrt2)^n  \negthinspace\negthinspace
\permsum{1}{n-1}
\sum_{j=1}^{n-2}
{
{ u^{\delta}(g)
  k_{j\delta\dot\gamma}
  \bar\kappa^{\dot\gamma\gamma}(j{+}1,n{-}1)
  u_{\gamma}(k_1) }
\over
{ \bra{g} 1,\ldots,n{-}1 \ket{1}  \braket{n}{g} }
}
\LF && \qquad\qquad\qquad
\phantom{\biggl[} \times\thinspace
{ \bar u_{\dot\alpha}(k_n)
 [\bar\kappa(j{+}1,n{-}1) - \bar\kappa(1,j{-}1)]^{\dot\alpha\alpha}
u_{\alpha}(g) }
\label{split}
\eeqa
We take advantage of the permutation sum to cyclicly relabel
the momenta in the second term as follows:
\beq
1 \rightarrow n{-}1 \rightarrow n{-}2 \rightarrow \cdots
\rightarrow 2 \rightarrow 1
\eeq
This converts the second term into
\beqa
\amp_2 = &&
{
{i}
\over
{48\pi^2}
}
(-e\sqrt2)^n  \negthinspace\negthinspace
\permsum{1}{n-1}
{
{ u^{\delta}(g)
  (k_{n-1})_{\delta\dot\gamma}
  \bar\kappa^{\dot\gamma\gamma}(1,n{-}2)
  u_{\gamma}(k_{n-1}) }
\over
{ \bra{g} n{-}1,1,2,\ldots,n{-}2 \ket{n{-}1}  \braket{n}{g} }
}
{ \bar u_{\dot\alpha}(k_n)
 \bar\kappa^{\dot\alpha\alpha}(1,n{-}2)
u_{\alpha}(g) }
\LF && +
{
{i}
\over
{48\pi^2}
}
(-e\sqrt2)^n  \negthinspace\negthinspace
\permsum{1}{n-1}
\sum_{j=2}^{n-2}
{
{ u^{\delta}(g)
  (k_{j-1})_{\delta\dot\gamma}
  \bar\kappa^{\dot\gamma\gamma}(j,n{-}2)
  u_{\gamma}(k_{n-1}) }
\over
{ \bra{g} n{-}1,1,2,\ldots,n{-}2 \ket{n{-}1}  \braket{n}{g} }
}
\LF && \qquad\qquad\qquad
\phantom{\biggl[} \times\thinspace
{ \bar u_{\dot\alpha}(k_n)
 [\bar\kappa(j,n{-}2)
    - \bar k_{n-1} -\bar\kappa(1,j{-}2)]^{\dot\alpha\alpha}
u_{\alpha}(g) }.
\label{twosplit}
\eeqa
A little bit of algebra allows us to rewrite this as
\beqa
\amp_2 = & - &
{
{i}
\over
{48\pi^2}
}
(-e\sqrt2)^n  \negthinspace\negthinspace
\permsum{1}{n-1}
\sum_{j=1}^{n-3}
{
{ u^{\delta}(g)
  k_{j\delta\dot\gamma}
  \bar\kappa^{\dot\gamma\gamma}(j{+}1,n{-}2)
  u_{\gamma}(k_{n-1}) }
\over
{ \bra{n{-}1} 1,\ldots,n{-}1 \ket{g}  \braket{n}{g} }
}
\LF && \qquad\qquad\qquad
\phantom{\biggl[} \times\thinspace
{ \bar u_{\dot\alpha}(k_n)
 [\bar\kappa(j{+}1,n{-}1)
    -\bar\kappa(1,j{-}1)]^{\dot\alpha\alpha}
u_{\alpha}(g) }
\LF & + &
{
{i}
\over
{48\pi^2}
}
(-e\sqrt2)^n \negthinspace\negthinspace
\permsum{1}{n-1}
\sum_{j=1}^{n-3}
{
{ u^{\delta}(g)
  k_{j\delta\dot\gamma}
  \bar\kappa^{\dot\gamma\gamma}(j{+}1,n{-}2)
  u_{\gamma}(k_{n-1}) }
\over
{ \bra{n{-}1} 1,\ldots,n{-}1 \ket{g}  \braket{n}{g} }
}
{ \bar u_{\dot\alpha}(k_n)
 [2\bar k_{n-1}]^{\dot\alpha\alpha}
u_{\alpha}(g) }
\LF & - &
{
{i}
\over
{48\pi^2}
}
(-e\sqrt2)^n  \negthinspace\negthinspace
\permsum{1}{n-1}
{
{ 2k_{n-1}\cdot k_n
  \braket{g}{n{-}1}
  {\braket{n{-}1}{n}}^{*} }
\over
{ \bra{n{-}1} 1,2,\ldots,n{-}2 \ket{n{-}1} \braket{n}{g} }
}.
\label{goforit}
\eeqa
The first term of~(\ref{goforit}) precisely cancels
$\amp_1$ [{\it i.e.}\ the first term of~(\ref{split})].
We apply the useful identity \cite{Mangano}
\beq
\permsum{1}{m}
{
{1}
\over
{ \bra{p}1,\ldots,m \ket{q} }
}
=
{
{ {\braket{p}{q}}^{m-1} }
\over
{ \prod\limits_{i=1}^{m} \bra{p} i \ket{q} }
}.
\label{permsummed}
\eeq
to see that the last term vanishes for $n>3$.
(We already know that the amplitude vanishes for $n\le3$, and so
lose no information by assuming that $n>3$.)  Consequently,
the entire amplitude may be written as
\beq
\amp(1^{+},\ldots,n^{+}) =  -
{
{i}
\over
{24\pi^2}
}
(-e\sqrt2)^n  \negthinspace\negthinspace
\permsum{1}{n-1}
\sum_{j=1}^{n-3}
{
{ u^{\delta}(g)
  k_{j\delta\dot\gamma}
  \bar\kappa^{\dot\gamma\gamma}(j{+}1,n{-}2)
  (k_{n-1})_{\gamma\dot\beta}
  \bar u^{\dot\beta}(k_n) }
\over
{ \bra{1} 2,\ldots,n{-}1 \ket{1}  \braket{n}{g} }
}
\label{remains}
\eeq
for $n\ge 4$.

In order to eliminate the last occurrence of the gauge momentum $g$,
it is necessary to re-order the terms a bit.  To this end, we
introduce the identity
\beq
\bar\kappa^{\dot\gamma\gamma}(j{+}1,n{-}2) u_{\gamma}(k_{n-1})
= \sum_{\ell=j+1}^{n-2}
\bar\kappa^{\dot\gamma\gamma}(j{+}1,\ell)u_{\gamma}(k_j)
\braket{j}{n{-}1} \link{\ell}{j}{\ell{+}1}.
\label{weirdID}
\eeq
Eq.~(\ref{weirdID}) is most easily proven by
performing the sum on $\ell$ appearing on the right hand side,
noting that
\beq
\link{1}{P}{2} + \link{2}{P}{3} = \link{1}{P}{3},
\eeq
a result that follows from the Schouten identity.  Application
of~(\ref{weirdID}) to~(\ref{remains}) produces
\beqa
\amp(1^{+}&&,\ldots,n^{+}) =
\LF && -
{
{i}
\over
{24\pi^2}
}
(-e\sqrt2)^n  \negthinspace\negthinspace
\permsum{1}{n-1}
\sum_{j=1}^{n-3}
\sum_{\ell=j+1}^{n-2}
{
{ u^{\delta}(g)
  k_{j\delta\dot\gamma}
  \bar\kappa^{\dot\gamma\gamma}(j{+}1,\ell)
  u_{\gamma}(k_j)
  u^{\beta}(k_j)
  (k_{n-1})_{\beta\dot\beta}
  \bar u^{\dot\beta}(k_n) }
\over
{ \bra{1} 2,\ldots\ell \ket{j}
  \bra{j} \ell{+}1,\ldots,n{-}1 \ket{1}  \braket{n}{g} }
}.
\label{uuu}
\eeqa
Let us examine the denominator of~(\ref{uuu}), which may
be written
\beq
\Delta = \bra{j} \ell{+}1, \ell{+}2, \ldots,
         n{-}1, 1, 2, \ldots, \ell \ket{j} \braket{n}{g}.
\label{string}
\eeq
Since $j<\ell$, we may express~(\ref{string}) in the form
\beq
\Delta = \bra{j} \ell{+}1, \ell{+}2, \ldots,
         n{-}1, 1, 2, \ldots, j{-}1 \ket{j}
         \bra{j} j{+}1, \ldots, \ell \ket{j}
         \braket{n}{g}.
\label{collapseform}
\eeq
Because the numerator of~(\ref{uuu}) is symmetric under permutations
of $\{j{+}1,\ldots,\ell\}$, we may use~(\ref{permsummed}) to
deduce that the permutation sum causes every term in the
sum over $\ell$ to vanish except for $\ell = j{+}1$.  Thus,
\beqa
\amp(1^{+}&&,\ldots,n^{+}) =
\LF && -
{
{i}
\over
{24\pi^2}
}
(-e\sqrt2)^n  \negthinspace\negthinspace
\permsum{1}{n-1}
\sum_{j=1}^{n-3}
{
{ u^{\delta}(g)
  k_{j\delta\dot\gamma}
  \bar k^{\dot\gamma\gamma}_{j+1}
  u_{\gamma}(k_j)
  \thinspace\thinspace
  \bar u_{\dot\beta}(k_n) }
  \bar k_{n-1}^{\dot\beta\beta}
  u_{\beta}(k_j)
\over
{ \bra{j} j{+}2, j{+}3,\ldots,n{-}1, 1, 2, \ldots j{-}1 \ket{j}
   \bra{j} j{+}1 \ket{j}  \braket{n}{g} }
}
\LF = && -
{
{i}
\over
{24\pi^2}
}
(-e\sqrt2)^n  \negthinspace\negthinspace
\permsum{1}{n-1}
\sum_{j=1}^{n-3}
{
{ u^{\delta}(g)
  k_{j\delta\dot\gamma}
  \bar k^{\dot\gamma\gamma}_{j+1}
  u_{\gamma}(k_{j+2})
  \thinspace\thinspace
  \bar u_{\dot\beta}(k_n) }
  \bar k_{n-1}^{\dot\beta\beta}
  u_{\beta}(k_j)
\over
{ \braket{j}{j{+}2}
  \bra{1} 2, 3,\ldots,n{-}1 \ket{1}
  \braket{n}{g} }
},
\label{vvv}
\eeqa
where we have done some minor rearranging to obtain the last line.

The denominator appearing in~(\ref{vvv}) is very nearly symmetric
under cyclic permutations of $\{1,2,\ldots,n{-}1\}$.
Let us take advantage of this near-cyclic symmetry to relabel the
successive terms appearing in the sum over $j$ so that
$k_j$ always becomes $k_1$, $k_{j+1}$ becomes
$k_2$, and $k_{j+2}$ becomes $k_3$.
The only portion of~(\ref{vvv}) which has
a form which varies after such relabeling
is $\bar k_{n-1}^{\dot\beta\beta}$.
In the $j=1$ term, $\bar k_{n-1}^{\dot\beta\beta}$ remains
unchanged.  For $j=2$, $\bar k_{n-1}^{\dot\beta\beta}$ becomes
$\bar k_{n-2}^{\dot\beta\beta}$, and so on through $j=n-3$,
in which $\bar k_{n-1}^{\dot\beta\beta}$ becomes
$\bar k_3^{\dot\beta\beta}$.  Hence, we see that~(\ref{vvv}) is
equivalent to
\beq
\amp(1^{+},\ldots,n^{+}) =
-
{
{i}
\over
{24\pi^2}
}
(-e\sqrt2)^n \negthinspace\negthinspace
\permsum{1}{n-1}
{
{ u^{\delta}(g)
  k_{1\delta\dot\gamma}
  \bar k^{\dot\gamma\gamma}_2
  u_{\gamma}(k_3)
  \thinspace\thinspace
  \bar u_{\dot\beta}(k_n) }
  \bar\kappa^{\dot\beta\beta}(3,n{-}1)
  u_{\beta}(k_1)
\over
{ \braket{1}{3}
  \bra{1} 2, 3,\ldots,n{-}1 \ket{1}
  \braket{n}{g} }
}.
\eeq
Momentum conservation for the scattering amplitude implies that
\beq
\kappa(3,n{-}1) = -k_1 - k_2 - k_n.
\eeq
Utilizing this relation plus a little more spinor algebra yields
\beq
\amp(1^{+},\ldots,n^{+}) =
{
{i}
\over
{24\pi^2}
}
(-e\sqrt2)^n  \negthinspace\negthinspace
\permsum{1}{n-1}
{
{ \braket{g}{1} {\braket{2}{1}}^{*} {\braket{n}{2}}^{*} }
\over
{ \bra{1} 3,\ldots,n{-}1 \ket{1} \braket{n}{g} }
}.
\label{nearlydone}
\eeq
Because the numerator of~(\ref{nearlydone}) is symmetric under
permutations of $\{3,\ldots,n{-}1\}$, we may conclude
from~(\ref{permsummed}) that the permutation sum will cause
the amplitude to vanish unless $n=4$:
\beq
\amp\bigl(1^{+},\ldots,n^{+}\bigr) =
0, \qquad n\ge5.
\label{photonphotonplus}
\eeq
Thus, we obtain the
extraordinary result that the {\it only}\ non-vanishing $n$-photon
amplitude in the case of like helicities is the set of $n=4$
box diagrams. We will comment further on this result in
section \ref{latersection}.

Although we have finished the derivation for $n>4$, we still
have not proven that the amplitude is independent of $g$ when
$n=4$.  We shall now do so.  For $n=4$, Eq.~(\ref{nearlydone})
reads
\beq
\amp(1^{+},2^{+},3^{+},4^{+}) =
{ {ie^4}\over{6\pi^2} }
\sum_{{\cal{P}}(123)}
{
{ \braket{g}{1} {\braket{2}{1}}^{*} {\braket{4}{2}}^{*} }
\over
{ \braket{1}{3} \braket{3}{1} \braket{4}{g} }
}.
\eeq
Since the denominator is symmetric under the interchange
$1\leftrightarrow3$,  we may easily
reduce the sum over all permutations
of $\{1,2,3\}$ to a sum over only cyclic permutations of
$\{1,2,3\}$ :
\beq
\amp(1^{+},2^{+},3^{+},4^{+}) =
{ {ie^4}\over{6\pi^2} }
\sum_{{\cal{C}}(123)}
{
{ {\braket{4}{2}}^{*} }
\over
{ \braket{1}{3} \braket{3}{1} \braket{4}{g} }
}
[ \braket{g}{1} {\braket{2}{1}}^{*}
 +\braket{g}{3} {\braket{2}{3}}^{*} ].
\label{lastg}
\eeq
Momentum conservation and the Weyl equation
allows us to rewrite
the combination in the square brackets  as
\beqa
u^{\alpha}(g)(k_1+k_3)_{\alpha\dot\alpha}\bar u^{\dot\alpha}(k_2)
&=& - u^{\alpha}(g)k_{4\alpha\dot\alpha} \bar u^{\dot\alpha}(k_2)
\LF
&=&  \braket{4}{g} \tekarb{2}{4}.
\label{thatdoesit}
\eeqa
Inserting~(\ref{thatdoesit}) into (\ref{lastg}) produces the
$g$-independent expression
\beq
\amp(1^{+},2^{+},3^{+},4^{+}) =
{ {ie^4}\over{6\pi^2} }
\sum_{{\cal{C}}(123)}
{
{ \tekarb{4}{2} \tekarb{2}{4} }
\over
{ \braket{1}{3} \braket{3}{1} }
}
\label{whew}
\eeq
Further judicious use of momentum conservation and the Weyl equation
reveals that the three terms of~(\ref{whew}) are actually
equal to each other.  Thus, with a bit more algebra,
we may write
\beq
\amp(1^{+},2^{+},3^{+},4^{+}) =
{ {ie^4}\over{2\pi^2} }
{
{ \tekarb{1}{2} \tekarb{3}{4} }
\over
{ \braket{1}{2} \braket{3}{4} }
}.
\label{fourphotonresult}
\eeq
In this form it is clear that the square of the amplitude
is constant
(up to an unobservable phase),
and agrees with the result for a massless fermion
loop
given by Gastmans and Wu \cite{ubiquitous}.

A similar calculation for
$\amp\bigl(1^{+},\ldots,(n-1)^{+},n^{-}\bigr)$
yields
\begin{mathletters}
\beq
\amp(1^{+},2^{+},3^{+},4^{-}) =
{ {ie^4}\over{2\pi^2} }
{
{ \tekarb{1}{2} \tekarb{2}{3} \braket{3}{1} }
\over
{ \braket{1}{2} \braket{2}{3} \tekarb{3}{1} }
},
\eeq
\beq
\amp\bigl(1^{+},\ldots,(n{-}1)^{+},n^{-}\bigr) =
0, \qquad n\ge5.
\label{photonphotonminus}
\eeq
\end{mathletters}

\subsection{Discussion}\label{latersection}

We now turn to a brief discussion of the surprising results
in Eqs.~(\ref{photonphotonplus}) and~(\ref{photonphotonminus}).
Our first observation is that we do not expect
{\it all}\
of the photon-photon helicity amplitudes to vanish for
$n\ge5$, unless $n$ is odd (Furry's theorem).
When there are two or more negative helicity photons present
in the diagram, it is possible to cut the diagram in such a
way so as to make the factors corresponding to
each piece non-zero.  Then, the Cutkosky rules imply the
existence of a non-vanishing imaginary part, unless some
special symmetry intervenes.  For example, in the case of
the photon-photon amplitudes for odd $n$,   charge conjugation
symmetry forces such a cancellation.  For even $n$, no such
symmetry is obvious.

Our second observation is that this result is confined to
the one-loop order of perturbation theory.  Indeed, it is easy
to see that the two-loop correction to
$\amp(1^{+},\ldots,6^{+})$ does not vanish.
The two types of diagrams contributing to this amplitude are
illustrated in Fig.~\ref{sixphoton}.  Of the possible diagrams
containing two fermion loops, only those which contain
three photons on each loop survive (Fig.~\ref{sixphoton}a).
Furry's theorem causes any diagram with an odd number of photons
attached to a loop to vanish.  Hence,
either the photons are divided equally,
as in Fig.~\ref{sixphoton}a,
or there is only one external photon on one of the loops and five
on the other.  But this latter arrangement consists of
the massless vacuum polarization
renormalization factor multiplied by the one-loop six-photon result:
this, of course, vanishes.
The diagrams like Fig.~\ref{sixphoton}a
may be viewed as a
pair of $n=4$ photon-photon scattering diagrams with one off-shell
photon.  These diagrams do not vanish when {\it all}
of the photons are on shell;  taking one of the photons
off shell cannot change this situation.
Thus, this contribution to
the amplitude is non-zero.
Furthermore, it should have a pole in the limit where the
photon connecting the two loops becomes soft.
The only other available diagrams are variants
of Fig.~\ref{sixphoton}b.  It is apparent that these
diagrams do not have the same pole structure.  Consequently,
they can not cancel the contribution from Fig.~\ref{sixphoton}a.
Hence, we expect a non-vanishing
result for  the two-loop six-photon amplitude with like
helicity photons.

\section{One-loop corrections to
$\eee^{+} \eee^{-} \rightarrow \gamma\gamma\cdots\gamma$}

We now turn to electron-positron annihilation into photons.
The one loop corrections to $e^{+} e^{-} \rightarrow \gamma\gamma$
have been known for some time \cite{BrownEtAl},
and
helicity amplitudes for the related process
of Compton scattering  are present
in the literature \cite{COMPTONloop} for full QED including
finite mass effects. Helicity
amplitudes for the massless limit, however,  seem to be
absent.  We will present results for an arbitrary number of
like-helicity photons in this limit.

\subsection{Preliminaries}

There are two basic types of diagrams contributing to the
one loop corrections to electron-positron
annihilation to $n$ photons,
as illustrated in Figs.~\ref{litefig} and~\ref{jellyfishfig}.
We will refer to Fig.~\ref{litefig} as  the light-by-light
contribution, since
it contains the light-by-light scattering
process as a sub-diagram, with one of the photons
off shell.   We will refer to Fig.~\ref{jellyfishfig}  as the
``jellyfish'' contribution since the off-shell photon in
Fig.~\ref{jellyfishfig} spans varying numbers of on-shell
photons.  The evaluation  of these two diagrams is relatively
easy.

\subsection{The light-by-light contribution}

Figure~\ref{litefig} illustrates the basic structure of
those Feynman diagrams which contain the light-by-light
scattering process as a sub-diagram.
The expressions relevant to this contribution may be generated
from~(\ref{lamdef}) by making the replacement
\beq
\eps_{\nu}(n) \rightarrow
-ie\bar\psi(p;1,\ldots,i) \gamma_{\nu}
\psi(i{+}1,\ldots,j;q)
\label{replacement}
\eeq
and introducing the appropriate sums to include all possible
divisions of the $n$ photons into three groups as well as
all possible permutations of the $n$ photons.  That is,
equation~(\ref{lamdef}) becomes
\vfill\eject
\beqa
\Lambda&&(p;1,\ldots,n;q) =
\LF && \enspace
-e^2 \negthinspace\negthinspace
\permsum{1}{n}
\sum_{j=0}^n
\sum_{i=0}^j
{
{1}
\over
{ i!(j{-}i)!(n{-}j)! }
}
\bar\psi(p;1,\ldots,i)\gamma_{\nu}
\psi(i{+}1,\ldots,j;q)
{\rm Tr} \thinspace
[ \gamma^{\nu} \Psi(\Poff;j{+}1,\ldots,n) ].
\LF &&
\label{litestart}
\eeqa
In equation~(\ref{litestart}),  $\bar\psi$
and $\psi$ stand for fermion currents with only
one end of the fermion line off shell.  Solutions for
these objects are discussed in
Refs.~\cite{BG} and~\cite{firstpaper}.
The incoming positron has momentum $p$, while the incoming
electron has momentum $q$.  The incoming photons have momenta
$k_1,\ldots,k_n$ as usual.
This expression is
valid for all combinations of
photon and fermion helicity.

Since the known double-off-shell fermion current contains
only like helicity photons, there are two helicity amplitudes
for this process which are readily accessible, corresponding to
the two possible
helicities of the fermion line.
For concreteness, we will discuss the amplitude containing
a left-handed positron.  The computation of the other amplitude
follows the same pattern. Alternatively,
one may apply charge-conjugation symmetry
to obtain the result.

Specializing then, to the left-handed positron case, and
converting~(\ref{litestart}) to spinor form produces
\beqa
\Lambda&&(p^{-};1,\ldots,n;q^{+}) =
\LF &&  \enspace
-(-e\sqrt2)^2 \negthinspace\negthinspace
\permsum{1}{n}
\sum_{j=0}^n
\sum_{i=0}^j
{
{1}
\over
{ i!(j{-}i)!(n{-}j)! }
}
\bar\psi^{\alpha}(p^{-};1,\ldots,i)(\sigma_{\nu})_{\alpha\dot\alpha}
\psi^{\dot\alpha}(i{+}1,\ldots,j;q^{+})
\LF && \qquad\qquad\qquad\qquad\qquad\times
{\rm Tr} \thinspace
[ (\sigma^{\nu})_{\beta\dot\beta}
  \bar\Psi^{\dot\beta\beta}(\Poff;j{+}1,\ldots,n)
 +(\bar\sigma^{\nu})^{\dot\beta\beta}
  \Psi_{\beta\dot\beta}(\Poff;j{+}1,\ldots,n) ].
\label{liteone}
\eeqa
Recall that the solutions for $\Psi$ and $\bar\Psi$ presented
in Eqs.~(\ref{LHsolution}) and~(\ref{RHsolution}) require that
a factor of the gauge spinor be contracted into the undotted
index.  The sum on $\nu$  appearing in~(\ref{liteone})
will cause the current $\bar\psi$ to appear in this position.
{}From Refs.~\cite{BG} and~\cite{firstpaper} we know that
\beq
\bar\psi^{\alpha}(p^{-};1^{+},\ldots,i^{+}) =
(-e\sqrt2)^i
\negthinspace\negthinspace
\permsum{1}{i}
{
{ u^{\alpha}(p) \braket{p}{g} }
\over
{ \bra{p} 1,\ldots,i \ket{g} }
}.
\label{psibar}
\eeq
Thus, the natural choice for the gauge momentum is $g=p$.
The expression in~(\ref{psibar}) vanishes for $i\ne0$
when $g=p$, and reads
\beq
\bar\psi^{\alpha}(p^{-}) =
 u^{\alpha}(p)
\eeq
for $i=0$.

The remaining steps in evaluating the contribution
from~(\ref{liteone}) involve straightforward spinor
algebra and an integration like that described in
Sec.~\ref{photonphoton}.  Therefore, we will immediately
present the result:
\beq
\amp_1(p^{-};1^{+},\ldots,n^{+};q^{+}) =
{
{-i}
\over
{24\pi^2}
}
(-e\sqrt2)^{n{+}2}
\permsum{1}{n}
\negthinspace\negthinspace
{
{ u^{\alpha}(p)
  k_{1\alpha\dot\alpha}
  {\bar{k}}_2^{\dot\alpha\beta}
  k_{3\beta\dot\beta}
  \bar\kappa^{\dot\beta\gamma}(1,3)
  u_{\gamma}(p) }
\over
{ \bra{1} 2,3 \ket{1} \thinspace
  \bra{p} 4, \ldots, n \ket{q} \thinspace
  \kappa^2(1,3) }
}.
\label{liteLHpositron}
\eeq
The structure of Eq.~(\ref{liteLHpositron}) reflects
the fact that the results obtained for photon-photon
scattering [in particular, Eq.~(\ref{photonphotonplus})]
continue to hold even when one of the photons has a non-zero
mass-squared [Note that the structure on the right-hand side
of~(\ref{replacement}) is transverse in the gauge utilized
in this calculation].  Thus, the only diagrams which actually
contribute to $\amp_1$ have exactly three external photons
attached to the fermion loop.  For $e^{+}e^{-}$ annihilation
to two photons, $\amp_1 \equiv 0$.

The corresponding amplitude with a right-handed positron is
given by
\beq
\amp_1(p^{+};1^{+},\ldots,n^{+};q^{-}) =
{
{i}
\over
{24\pi^2}
}
(-e\sqrt2)^{n{+}2}
\permsum{1}{n}
\negthinspace\negthinspace
{
{ u^{\alpha}(q)
  k_{1\alpha\dot\alpha}
  {\bar{k}}_2^{\dot\alpha\beta}
  k_{3\beta\dot\beta}
  \bar\kappa^{\dot\beta\gamma}(1,3)
  u_{\gamma}(q) }
\over
{ \bra{1} 2,3 \ket{1} \thinspace
  \bra{p} 4, \ldots, n \ket{q} \thinspace
  \kappa^2(1,3) }
}.
\label{liteRHpositron}
\eeq

\subsection{The ``jellyfish'' contribution}

Figure~\ref{jellyfishfig} illustrates the basic structure of
this group of Feynman diagrams.  The direct application of
ordinary QED Feynman rules to produces
\beqa
\amp_2&&(p^{-};1,\ldots,n;q^{+}) =
\LF && \qquad
ie^2 \negthinspace\negthinspace
\permsum{1}{n}
\sum_{i=0}^{n}  \sum_{j=i}^n
{
{1}
\over
{ i! (j{-}i)! (n{-}j)!}
}
\bar \psi(p;1,\ldots,i) {\half}(1-\gamma^5) \gamma^{\nu}
\LF && \qquad\qquad\qquad\qquad\enspace\times
\Psi(\Poff;i{+}1,\ldots,j)
\gamma_{\nu} {\half}(1+\gamma^5)\psi(j{+}1,\ldots,n;q)
{
{1}
\over
{\Koff^2}
}
\label{startjelly}
\eeqa
where we have inserted
the appropriate projection
operator to describe
a left-handed positron.
The spanning photon in Fig.~\ref{jellyfishfig} has momentum
$\Koff$.
The momentum of the
off-shell positron in $\Psi$ is given by
\beq
\Poff \equiv \Koff + p + \kappa(1,i).
\eeq
We again choose $g=p$ to
ensure the
desired contraction of the gauge spinor into the undotted
index of $\Psi$ or $\bar\Psi$.
This also
reduces the three currents appearing
in~(\ref{startjelly}) to two.
Making this gauge selection, translating
to spinor notation, and performing the sum on $\nu$ yields
\beqa
\amp_2&&(p^{-};1^{+},\ldots,n^{+};q^{+}) =
\LF &&
-2ie^2 \negthinspace\negthinspace
\permsum{1}{n}
\sum_{j=0}^n
{
{1}
\over
{ j!(n{-}j)! }
}
\psi_{\dot\alpha}\bigl( (j{+}1)^{+}, \ldots, n^{+};q^{+} \bigr)
\Psibar^{\dot\alpha\alpha} (\Poff;1^{+},\ldots,j^{+})
u_{\alpha}(p)
{
{1}
\over
{\Koff}^2
}.
\label{jellyone}
\eeqa

The $j=0$ term in Eq.~(\ref{jellyone}) is special:  it contains
the fermion self-energy as a subgraph.
The contribution from this term is proportional to the tree-level
result.  Since the tree-level diagram vanishes
for this helicity combination,
we may drop $j=0$ from the sum in (\ref{jellyone}).
We now insert the expressions for the currents in the remaining
terms of (\ref{jellyone}) to obtain
\beqa
\amp_2&&(p^{-};1^{+},\ldots,n^{+};q^{+}) =
\LF && \quad
-(-e\sqrt2)^{n+2}
\permsum{1}{n}
\sum_{j=1}^n \sum_{\ell=1}^j
\int
\dfour{\Koff}
{
{ u^{\beta}(p)
  [ \kappa(j{+}1,n){+}q ]_{\beta\dot\alpha}
  \bar\Koff^{\dot\alpha\alpha}
  u_{\alpha}(p) }
\over
{ \bra{p}1,\ldots,j \ket{p}
  \bra{p}j{+}1,\ldots,n\ket{q} }
}
\LF && \qquad\qquad\qquad\qquad \times
{
{ u^{\gamma}(p)
  k_{\ell\gamma\dot\gamma}
  [ \bar\Koff{+}\bar p{+}\bar\kappa(1,\ell)]^{\dot\gamma\delta}
  u_{\delta}(p) }
\over
{ \Koff^2 [\Koff{+}p{+}\kappa(1,\ell{-}1)]^2
          [\Koff{+}p{+}\kappa(1,\ell)]^2 }
}.
\label{jellytwo}
\eeqa
While it is possible to do the integral at this stage, it
is  advantageous to simplify the integrand as much
as possible first.  To this end, we perform the sum on
$j$, producing
\beqa
\amp_2&&(p^{-};1^{+},\ldots,n^{+};q^{+}) =
\LF &&
(-e\sqrt2)^{n+2}
\permsum{1}{n}
\sum_{\ell=1}^n
\int
\dfour{\Koff}
{
{ u^{\alpha}(p)
  \Koff_{\alpha\dot\alpha}
  [\bar p {+} \bar\kappa(1,\ell{-}1)]^{\dot\alpha\beta}
  k_{\ell\beta\dot\gamma}
  [ \bar\Koff{+}\bar p{+}\bar\kappa(1,\ell)]^{\dot\gamma\gamma}
  u_{\gamma}(p) }
\over
{ \bra{p}1,\ldots,n \ket{q}
  \Koff^2 [\Koff{+}p{+}\kappa(1,\ell{-}1)]^2
          [\Koff{+}p{+}\kappa(1,\ell)]^2 }
}.
\label{jellythree}
\eeqa
Next, we would like to write
\beqa
p+\kappa(1,\ell) = [\Koff + p + \kappa(1,\ell)]
          -\Koff
\eeqa
in order to begin the process of canceling some of the denominators.
However, this would break up a convergent integral into divergent
bits.  We must regulate this expression first, as discussed in
the Sec.~\ref{INTEGRATION}.  As was true in that
case, every occurrence of the loop momentum may be written as
a Lorentz dot product with one of the polarization vectors.
Thus, Eq.~(\ref{jellythree}) may be continued to $d$ dimensions
without difficulty.  The subsequent
algebra to actually cancel the denominators
is very similar to that performed in
Eqs.~(\ref{numdef})--(\ref{goodstuff}).
Hence, we shall immediately quote the result:
\beqa
\amp_2&&(p^{-};1^{+},\ldots,n^{+};q^{+}) =
\LF &&
-(-e\sqrt2)^{n+2} \negthinspace
\permsum{1}{n}
\sum_{\ell=1}^{n+1}
{
{1}
\over
{ \bra{p}1,\ldots,n\ket{q} }
}
\LF && \quad\qquad\qquad\qquad\qquad\times
\int
\dn{\Koff}
{
{ u^{\alpha}(p)
  k_{\ell\alpha\dot\alpha}
  [ \bar K{+}\bar p{+}\bar\kappa(1,\ell)]^{\dot\alpha\gamma}
  u_{\gamma}(p) }
\over
{  \{ [K{+}p{+}\kappa(1,\ell{-}1)]^2-\mu^2 \}
   \{ [K{+}p{+}\kappa(1,\ell)]^2 -\mu^2 \} }
}
\LF &&
+(-e\sqrt2)^{n+2} \negthinspace
\permsum{1}{n}
\sum_{\ell=2}^{n}
{
{1}
\over
{ \bra{p}1,\ldots,n\ket{q} }
}
\int
\dn{\Koff}
{
{1}
\over
{ [K^2 - \mu^2] }
}
\LF && \quad\qquad\qquad\qquad\qquad\times
{
{ \mu^2\thinspace u^{\alpha}(p)
  k_{\ell\alpha\dot\alpha}
  [ \bar p{+}\bar\kappa(1,\ell)]^{\dot\alpha\gamma}
  u_{\gamma}(p) }
\over
{  \{ [K{+}p{+}\kappa(1,\ell{-}1)]^2 - \mu^2 \}
   \{ [K{+}p{+}\kappa(1,\ell)]^2-\mu^2 \} }
}
\label{jellyfour}
\eeqa
where $k_{n+1}\equiv q$.
The  first integral appearing
in~(\ref{jellyfour}) contains the same denominator structure
as the integral in~(\ref{zerointegral}).  Since
the numerator of this term
contains the same factor that caused~(\ref{zerointegral})
to vanish
[see Eq.~(\ref{zeroshift})],
this integral vanishes as well.  The second integral
may be evaluated with little difficulty, yielding
\beq
\amp_2(p^{-};1^{+},\ldots,n^{+};q^{+}) =
{
{-i}
\over
{32\pi^2}
}
(-e\sqrt2)^{n+2} \negthinspace\negthinspace
\permsum{1}{n}
\sum_{\ell=1}^{n}
{
{ u^{\alpha}(p)
  k_{\ell\alpha\dot\alpha}
  [\bar p +\bar\kappa(1,\ell)]^{\dot\alpha\gamma}
  u_{\gamma}(p) }
\over
{ \bra{p}1,\ldots,n\ket{q} }
}.
\label{jellyLHdone}
\eeq
A similar calculation utilizing a right-handed positron
produces
\beq
\amp_2(p^{+};1^{+},\ldots,n^{+};q^{-}) =
{
{i}
\over
{32\pi^2}
}
(-e\sqrt2)^{n+2} \negthinspace\negthinspace
\permsum{1}{n}
\sum_{\ell=1}^{n}
{
{ u^{\alpha}(q)
  k_{\ell\alpha\dot\alpha}
  [\bar p +\bar\kappa(1,\ell)]^{\dot\alpha\gamma}
  u_{\gamma}(q) }
\over
{ \bra{p}1,\ldots,n\ket{q} }
}.
\label{jellyRHdone}
\eeq
The total one-loop amplitudes are, of course, given by the
sums of~(\ref{liteLHpositron}) and~(\ref{jellyLHdone})
or~(\ref{liteRHpositron}) and~(\ref{jellyRHdone}) depending
upon the helicity of the fermion line.

\section{Conclusions}

In this paper we have seen a simple way
to compute the one-loop corrections
to QED helicity amplitudes that vanish at tree level.
We know from analyticity considerations that the
expressions should be relatively simple:  in particular, there
should be no cuts.
By using the solutions to the
recursion relations for currents that contain
two off shell particles, it is possible to
evaluate these
amplitudes easily, without the production of extraneous
logarithmic contributions which cancel in the end.
The amount of labor involved is essentially independent of
the number of photons.

We have obtained helicity amplitudes for $n$-photon scattering
as well as electron-positron annihilation to $n$ photons for
the case of like helicity photons. We have also obtained
$n$-photon scattering amplitudes for the case where one
of the photon helicities is the opposite of the rest.
We find that the only non-vanishing $n$-photon scattering
amplitudes for these helicity combinations are for $n=4$.
This is a new and surprising result.

In principle, the extension of these methods to more
complicated helicity combinations is straightforward, although
somewhat more computational labor is required.
It is likely that the quest for
closed-form expressions for arbitrary $n$
will have to end, however.
Instead, one should focus on using the
recursion relations as a powerful guide in simplifying the
integrand as much as possible before attacking the actual
integration.  Indeed, the reductions illustrated here suggest
that except for the most complicated helicity configuration
({\it i.e.}\  equal numbers of positive and negative helicity
photons), there is much to be learned from this point of
view.

\acknowledgments

I would like to thank T.--M. Yan, C.--S. Lam,
R.T. Ward, B. Dimm, S. Parke, W.T.~Giele
and D. Summers for useful discussions during the course of this
work.
Additional thanks to T.--M.~Yan and W.T.~Giele for reading and
commenting on the manuscript prior to release.
I would also like to thank B. Dimm for
the use of FeynDiagram to generate the figures for this paper.

This work was started at Cornell University and supported in part
by the National Science Foundation.

\begin{figure}[h]

\caption[]{The basic diagram for $n$-photon scattering.  The blob
represents the sum of all possible tree graphs with $n-1$ photons
attached to a fermion line.}\label{ngammafig}

\caption[]{Two of the two-loop diagrams for
$\gamma\gamma \rightarrow \gamma\gamma\gamma\gamma$.
\hskip5.95cm}\label{sixphoton}

\caption[]{The light-by-light diagram for the process
$e^{+} e^{-} \rightarrow \gamma\gamma\ldots\gamma$.
\hskip5.95cm}\label{litefig}

\caption[]{The ``jellyfish''  diagram for the process
$e^{+} e^{-} \rightarrow \gamma\gamma\ldots\gamma$.
\hskip5.95cm}\label{jellyfishfig}

\end{figure}


\begin{references}
\bibitem{other}  Additional references to the
vast literature on this subject may be found within
Refs.~\cite{ubiquitous} and~\cite{review}.  \label{Other}
\end{references}
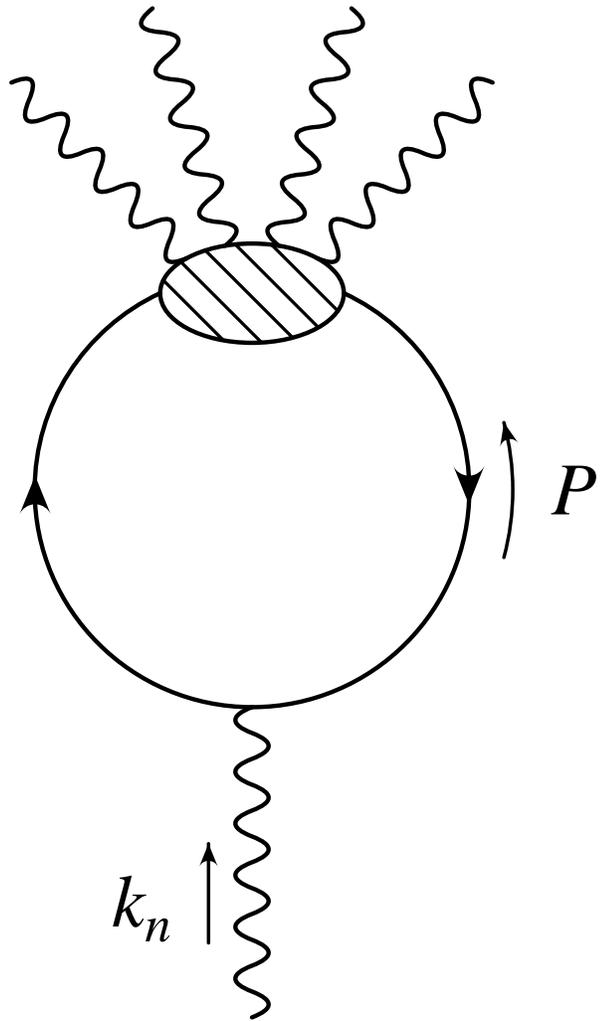
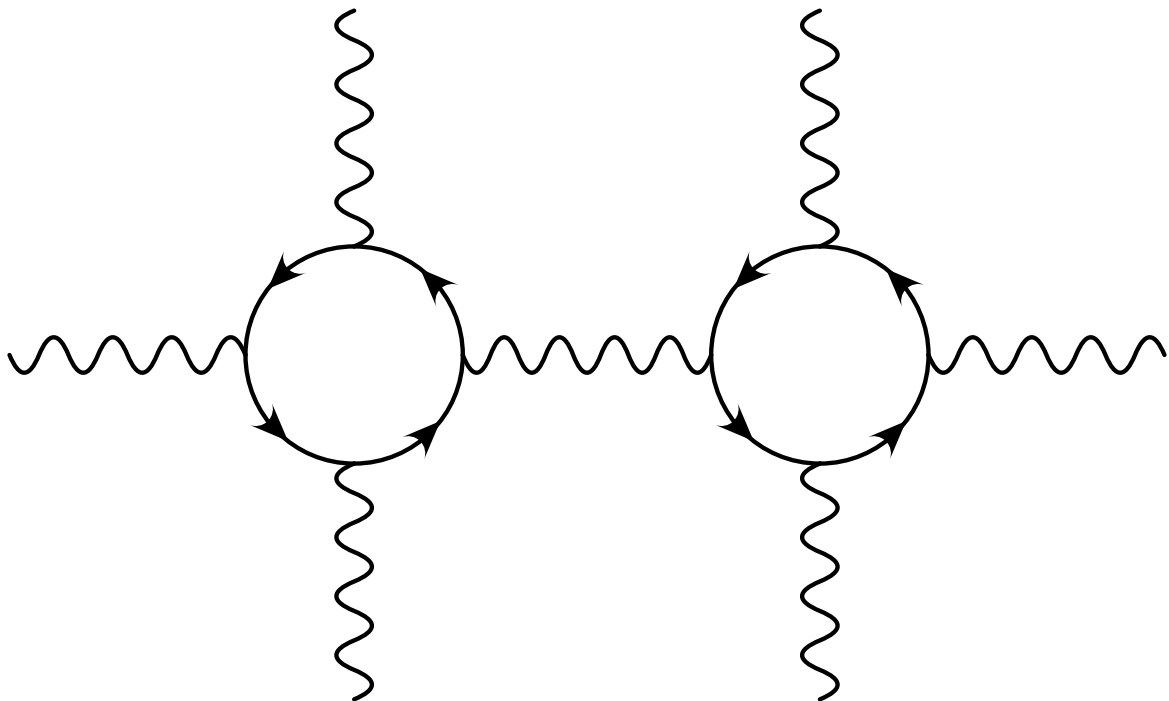
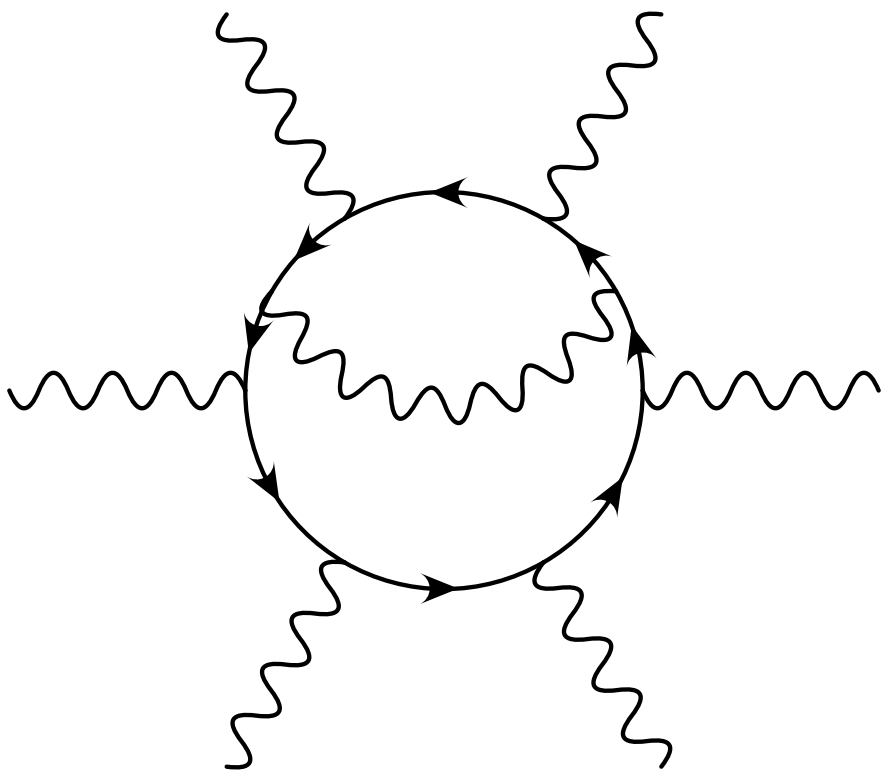
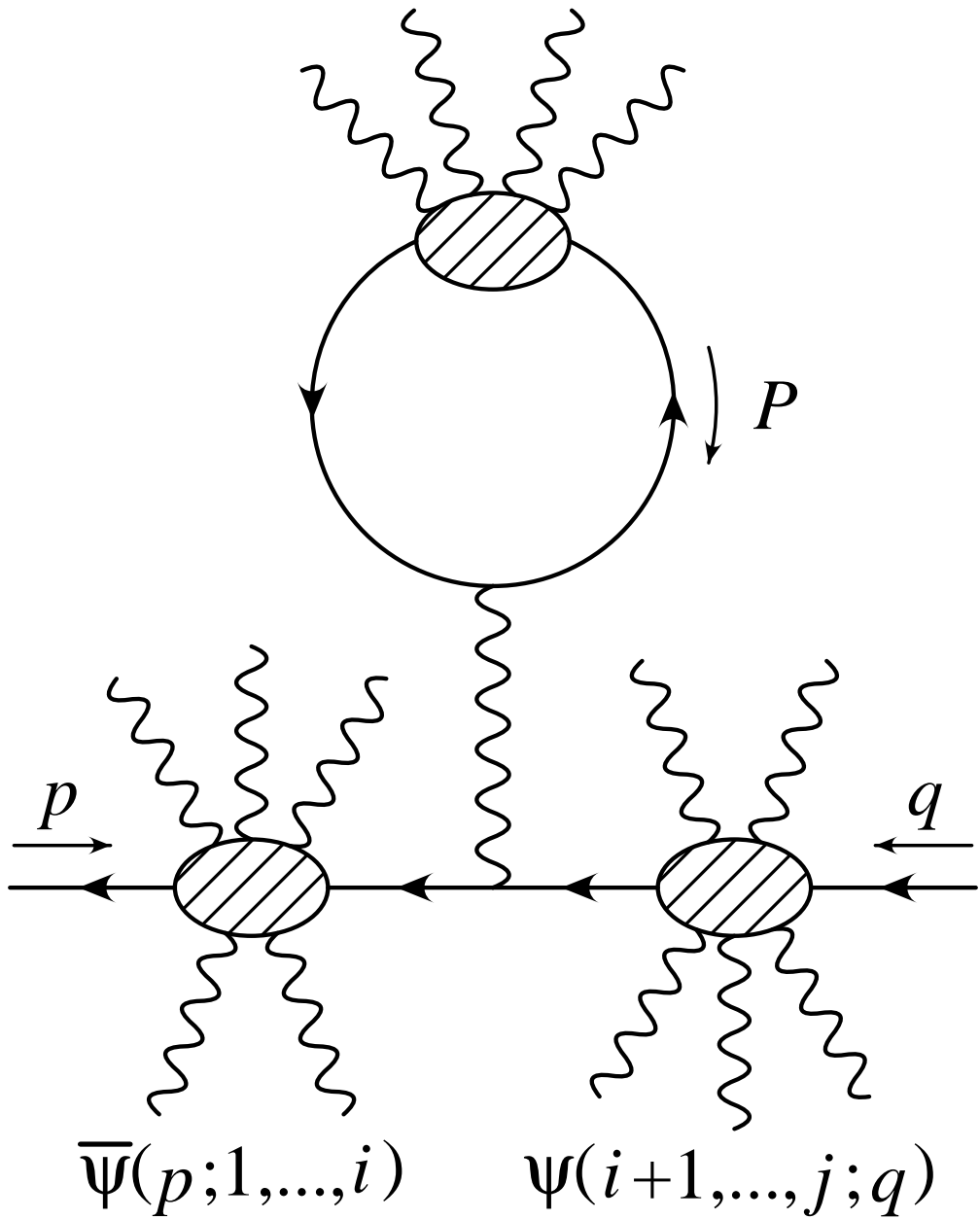
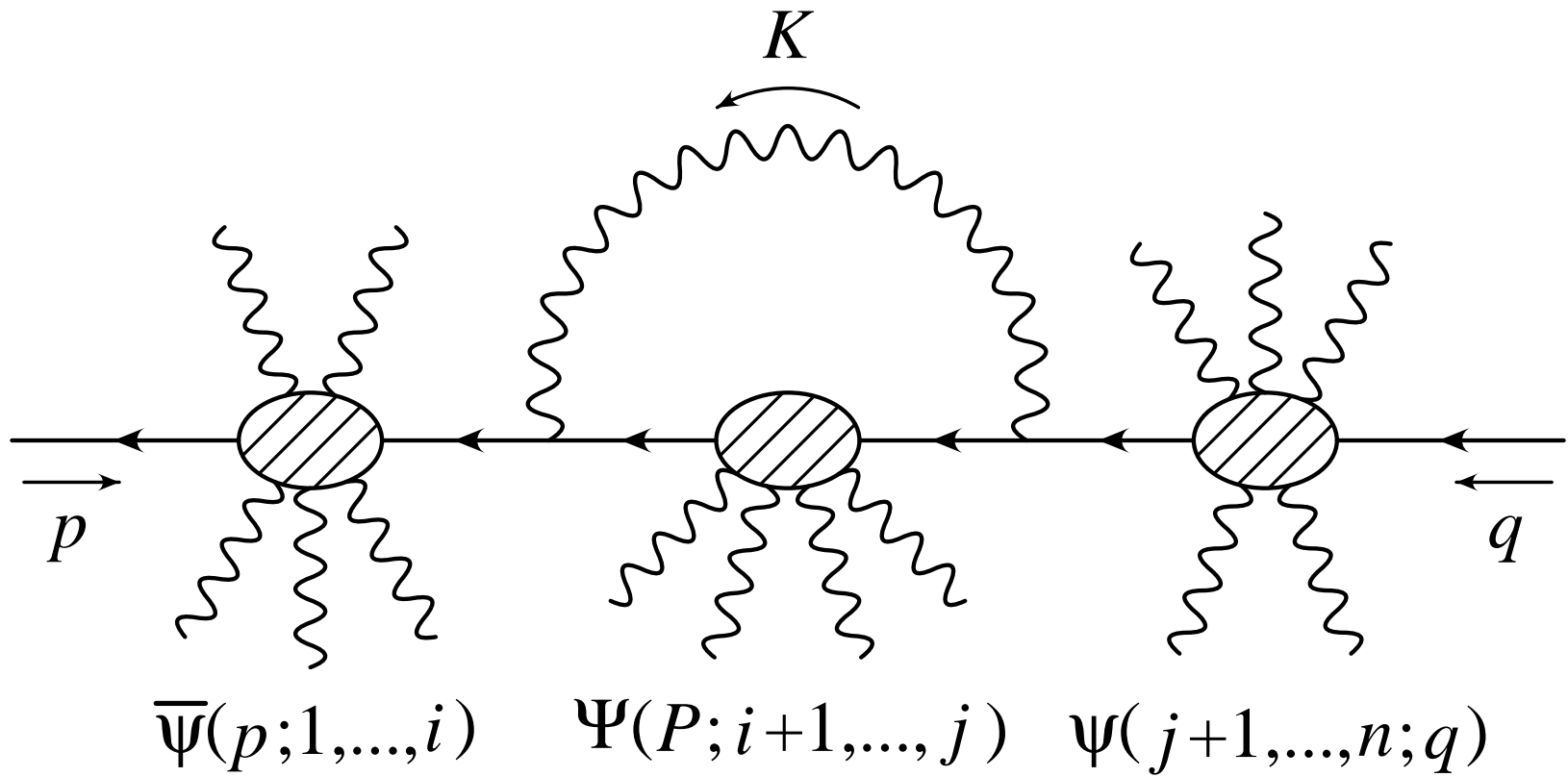
\end{document}